\documentclass[%
 reprint,
 amsmath,amssymb,
 aps,
]{revtex4-1}

\usepackage{graphicx}% Include figure files
\usepackage{dcolumn}% Align table columns on decimal point
\usepackage{bm}% bold math
\begin{document}

\preprint{APS/123-QED}

\title{Graphene-based three-body amplification of photon heat tunneling}

\author{Hamidreza Simchi}
\email{simchi@alumni.iust.ac.ir}
\affiliation {Department of Physics, Iran University of Science and Technology, Narmak, Tehran 16844, Iran} \affiliation{Semiconductor Technoloy Center, P.O.Box 19575-199, Tehran, Iran}%Lines break automatically or can be forced with \\

\date{\today}% It is always \today, today,

\begin{abstract}
We consider a three slabs configuration including two non-doped single layer graphene on insulating silicon dioxide (G/SiO$_2$) substrates and one non-doped suspended single layer graphene (SG). The suspended layer is placed between two G/SiO$_2$ layers. Without SG layer, the heat flux has maximum at Plasmon frequency supported by the G/SiO$_2$ slabs. In three slabs configuration, the photon heat tunneling is amplified between two G/SiO$_2$ layers significantly, only for specific range of vacuum gap between SG layer and G/SiO$_2$ layers and Plasmon frequency, due to the coupling of modes between each G/SiO$_2$ layer and SG layer. Since, the SG layer is a single atomic layer, the photon heat tunneling assisted by this configuration does not depend on the thickness of middle layer and in consequence, it can enable novel applications for nanoscale thermal management.
\end{abstract}

\maketitle

%\tableofcontents

\section{Introduction}
In any material, there will be spontaneous electrical and magnetic moments that originate from quantum and thermal fluctuations. These fluctuating moments produce fluctuating electromagnetic fields inside and outside the material. One can calculate these fluctuating fields by adding a fluctuating induction terms to the Maxwell's equation \cite{R1}. Lifshitz has found the fields inside and outside the materials by using the fluctuation-dissipation theorem \cite{R2}. Since 1930, London has realized that quantum mechanical fluctuations of electric dipole moments could give rise to the force between bodies separated by macroscopic distances (van der Waals force) \cite{R3}. Abrikosova  \textit{et al.} have made the first direct measurements of the van der Waals force between macroscopic bodies \cite{R4}.  The energy flow between two half spaces has been found using Lifshitz's method by using the Poynting vector rather than the Maxwell stress tensor \cite{R5}. Sipe has developed new Green function formalism for calculating fields generated by sources in the presence of a multilayer geometry \cite{R6}. Narayanaswamy  \textit{et al.} have found a relation between cross-spectral densities of electromagnetic fields in thermal equilibrium and dyadic Green functions of the vector Helmholtz equation \cite{R7} and then generalized it to thermal non-equilibrium effects (i.e., when the objects are at different temperatures) and introduced a Green function formalism of energy and momentum transfer in fluctuational electrodynamics \cite{R8}.\\  
But, what will be the force and heat transfer between bodies separated by nanometer distances? Several groups \cite{R5,R9,Rten,R11} have shown that when the gap distance, $d$ between bodies becomes very small, the near-field heat transfer varies as $d^{-2}$. By using nonlocal dielectric function it has been shown that, the $d^{-2}$ dependence would disappear for $d<0.1$ nanometer (nm)\cite{R12},\cite{R13} but generally speaking, the nonlocal effects has little influence on the predicted heat flux for $d>0.1$ nm\cite{R12}. Basu  \textit{et al.} have considered two semi-infinite plates separated by a vacuum gap of finite width, especially for $0.1<d<100$ nm, and studied the dependency of maximum heat flux to the dielectric function and vacuum gap width \cite{R14}. In addition at subwavelength gap (i.e., in near field regime), it has been shown that, a significant increment of heat flux results from the evanescent photons which remain confined near the surface of the materials \cite{R15,R16,R17,R18,R19,R20,R21,R22}. Messina et al., have considered a metal-like (ML) medium layer (with width, $\delta$) between two silicon carbide (SiC) layers (with width, $d$) and studied the amplification of photon heat tunneling in this structure \cite{R23}. They have shown that, only for $0.05<\delta<0.2$ micrometer $(\mu m)$ and $0.05<d<0.3$ $\mu m$, the amplification was occurred\cite{R23}. The amplification of photon heat has opened new possibilities for development some technologies such as, thermophotovoltaic conversion devices \cite{R24, R25}, Plasmon assisted nano-photolithography \cite{R26} and infrared sensing and spectroscopy \cite{R27,R28}.
\\
In the present work, we consider three-body configuration including a non-doped suspended single layer graphene (SG) which is placed between two non-doped single layer graphene on silicon dioxide (G/SiO$_2$) substrates. At first, we calculate the photon heat tunneling between two G/SiO$_2$ structure and show: how the heat flow depends on the value of gap ($d$) between these layers and frequency of filed ($\omega$). Then, we place the SG layer between these two G/SiO$_2$ structures, when the heat flux on SG layer is zero, and show that, the heat transfer between G/SiO$_2$ substrates is amplified only for specific range of vacuum gap between SG and each  G/SiO$_2$ substrate i.e, $d$ and Plasmon frquency. The structure of the article is as follows: in section II the analytical calculation method of dielectric constant of non-doped graphene and evanescent part of heat transfer will be presented. In section III the result and discussion and in section IV the summary will be provided, respectively.
\section{Analytical calculations}
In this section of article, we provide the calculation method of dielectric constant and evanescent part of heat transfer.
\subsection{Dielectric constant of undoped graphene}
The expectation value of an arbitrary observable $\hat A$ under the influence of a probe potential of the form:
\begin{equation}
\hat V(t)= \hat B F(t)\theta [t-t_0] 
\end{equation}                                                             
with $\hat B $ being an observable,$F(t)$ a scalar function of time, and $\theta (t)$ the Heaviside step function, can be written as:
\begin{equation}
{〈\delta\hat A (t)〉}=\int_{-\infty}^{+\infty}{\frac{d\omega}{2\pi}}{ \chi_{AB} (\omega)}{F(\omega)}{e^{-i\omega t}}
\end{equation}  
where $F(\omega)$ is Fourier transformation of $F(t)$  and $\chi_{AB }(\omega)$  is complex-valued generalized susceptibility, which is equal to:
\begin{equation}
{ \chi_{AB} (\omega)}={-i}\int_{0}^{+\infty}{d}{ t^\prime}{〈[\hat A(t^\prime),\hat B]〉}_{eq} {e^{(i\omega t^\prime-0t^\prime )}}
\end{equation}
This quantity can be written with a slightly different, more convenient, notation as \cite{R29, R30}:
\begin{widetext}
\begin{equation}
\chi(\vec{q},\omega)=4\sum_{\alpha,\alpha^\prime=\pm 1} \int{\frac{d^2\vec{k}}{4\pi^2}}{\frac{({n_F(\alpha E_k)} - {n_F(\alpha^\prime E_{k+q})})}{(\omega + \alpha E_k - \alpha^\prime E_{k+q} + i\epsilon)}} F_{\alpha \alpha^\prime} (\vec{q},\vec{k})
\label{eq:wideeq}
\end{equation}
\end{widetext}
where $\epsilon=0^+$, $\alpha(\alpha^\prime)$ index of band, and
\begin{equation}
F_{\alpha \alpha^\prime} (\vec{q},\vec{k})={\frac{1}{2}}(1+\alpha\alpha^\prime {\frac{k^2+\vec{k}.\vec{q}}{{k}\mid{\vec{k}+\vec{q}}\mid})}
\end{equation}
\begin{equation}
n_F (E)=[e^{\beta \hbar \omega}+1]^{-1}     
\end{equation}
where $\beta=(k_B T)^{-1}$ , with $T$ the temperature and $k_B$ the Boltzmann constant. If $\theta $ is angle between $\vec{k}$ and $\vec{q}$ then
\begin{equation}
F_{\alpha \alpha^\prime} (\vec{q},\vec{k})={\frac{1}{2}}(1+\alpha\alpha^\prime {\frac{k+q\cos(\theta)}{\mid{\vec{k}+\vec{q}}\mid})}
\end{equation}
By changing the variable, $u=(1+\alpha\alpha^\prime {\frac{k+q\cos(\theta)}{\mid{\vec{k}+\vec{q}}\mid}})$  and using the below relations: 
\begin{equation}
\omega-k=\mid{\vec{k}+\vec{q}}\mid
\end{equation}
\begin{equation}
\cos(\theta)={\frac{(k+q\cos(thetaθ))}{(\omega-k)}}
\end{equation}
\begin{equation}
\sin(\theta)=\pm{\frac{1}{2kq}}\sqrt{{4k^2 q^2-(\omega^2-2\omega k-q^2 )^2}}
\end{equation}
The imaginary part of $\chi(\vec{q},\omega)$ becomes (for $q<\omega$):
\begin{equation}
Im(\chi(\vec{q},\omega))={\frac{-1}{\pi\sqrt{\omega^2-q^2}}}\int_{\frac{\omega-q}{2}}^{\frac{\omega+q}{2}}{d}{ k}\sqrt{q^2-\omega^2+4k\omega-4k^2}
\end{equation}
But, $\sqrt{q^2-\omega^2+4k\omega-4k^2}=\sqrt{q^2-(\omega+2k)^2}$, and if $x\equiv\mid\vec{q}\mid/\mid\vec{k_F}\mid$ and $\upsilon=\omega/E_F$, it can be shown\cite{R30,R31}:
\begin{equation}
Im(\chi(\vec{q},\omega))=-{\frac{x^2\theta(\upsilon-x)}{4\sqrt{\upsilon^2-x^2}} },   q<\omega
\end{equation}
\begin{equation}
Im(\chi(\vec{q},\omega))=0,  q>\omega 
\end{equation}
Also, based on the Kramers-Kronig relation, the relation between $Im(\chi(\vec{q},\omega))$ and $Re(\chi(\vec{q},\omega))$ is as follows\cite{R30}:
\begin{equation}
Re(\chi(\vec{q},\omega))={\frac{2}{\pi}} P \int_{0}^{\infty}{d}{ \omega^\prime}{\frac {\omega^{\prime} Im(\chi(\vec{q},\omega)) }{\omega^{\prime^2} -\omega^2}}
\end{equation}
where, $P$ means principle vlaue. Therefore\cite{R30}:
\begin{equation}
Re(\chi(\vec{q},\omega))=-{\frac{x^2\theta(x-\upsilon)}{4\sqrt{x^2-\upsilon^2}} },    q>\omega
\end{equation}
\begin{equation}
Re(\chi(\vec{q},\omega))=0,    q<\omega
\end{equation}
However, based on the random phase approximation, the relation between dielectric constant $\varepsilon(\vec{q},\omega)$ and susceptibility $\chi(\vec{q},\omega)$  can be written as \cite{R31}:
\begin{equation}
\varepsilon(\vec{q},\omega)=1-V(q)\chi(\vec{q},\omega)
\end{equation}                                                        
where, $V(q)=2\pi e^2/\kappa q$ is the two dimensional Coulomb interaction. Therefore \cite{R31},
\begin{equation}
Re(\chi(\vec{q},\omega))=1+{\frac{e^2}{2\varepsilon_0\varepsilon_r}}(1+x\theta(x\pm\upsilon)/(4\sqrt{x^2-\upsilon^2}))
\end{equation}
\begin{equation}
Im(\chi(\vec{q},\omega))={\frac{e^2}{2\varepsilon_0\varepsilon_r}}x\theta(\pm\upsilon-x)/(4\sqrt{\upsilon^2-x^2})
\end{equation}
Now, if $\upsilon>x$ then:
\begin{equation}
Re(\chi(\vec{q},\omega))=1+{\frac{e^2}{2\varepsilon_0\varepsilon_r}}
\end{equation}
\begin{equation}
Im(\chi(\vec{q},\omega))={\frac{e^2}{2\varepsilon_0\varepsilon_r}}x/(4\sqrt{\upsilon^2-x^2})
\end{equation}
\subsection{Evanescent photon heat tunneling}
The three-body configuration is shown in Fig.1. First, we consider the structure without intermediate suspended graphene layer. We assume that the materials are nonmagnetic.
\begin{figure}[b]
\includegraphics{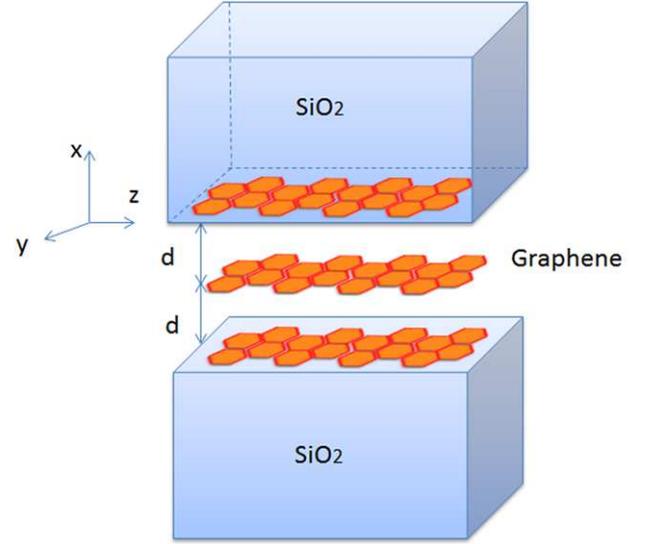}% Here is how to import EPS art
\caption{\label{fig:epsart} Schematic of three-body configuration for photon heat tunneling calculation.}
\end{figure}
 It is obvious that, the electric and magnetic field that are generated by the component of the fluctuating induction $\vec{g}(\vec{r},t)$ hold in the Maxwell's equations\cite{R5}
\begin{equation}
\nabla\times\vec{E}(\vec{r},\omega)=i{\frac{\omega}{c}}\vec{H}(\vec{r},\omega)
\end{equation}
\begin{equation}
\nabla\times\vec{H}(\vec{r},\omega)=-i{\frac{\omega}{c}}\vec{E}(\vec{r},\omega)-i{\frac{\omega}{c}}\vec{g}(\vec{r},t)
\end{equation}
We can write the fields in terms of Fourier components as:
\begin{widetext}
\begin{equation}
\vec{E}={\int_{-\infty}^{+\infty}}{\vec{a}}(\vec{k})e^{i\vec{k}.\vec{r}}{d}{\vec{k}} ,        \vec{H}={\frac{c}{\omega}}{\int_{-\infty}^{+\infty}}{\vec{k}}{\times}{\vec{a}}(\vec{k})e^{i\vec{k}.\vec{r}}{d}{\vec{k}}
\label{eq:wideeq}
\end{equation}
\end{widetext}
and $\vec{g}$ as:
\begin{equation}
\vec{g}={\int_{-\infty}^{+\infty}}{\vec{g}}(\vec{k})e^{i\vec{k}.\vec{r}}{d}{\vec{k}}
\end{equation}
By substituting Eqs.24 and 25 in Eqs.22 and 23, and defining the below vectors:
\begin{equation}
\vec{k}_{\pm}=\vec{q}+\pm s_1\hat x, \vec{q}=k_y\hat y+k_z\hat z, s_1=\sqrt{\omega^2\varepsilon_{SiO_{2}}-q^2} 
\end{equation}
\begin{equation}
\hat s=\hat q\times\hat x, \hat p_{\pm}=k^{-1}(q\hat x\mp s_1\hat q)
\end{equation}
\begin{equation}
\vec{\rho}=y\hat y+z\hat z
\end{equation}
The transverse and parallel component of filed become \cite{R5}:
\begin{widetext}
\begin{equation}
\vec{E}^{\perp}=\int{\frac{i\pi\omega^{SiO_{2}}{t_{10}^{\perp}}}{c^2q^2s_1}}exp[i\vec{q}.\vec{\rho}+ipx]\times[\vec{g}(\vec{k}).(\vec{q}\times\hat x)](\vec{q}\times\hat x)d^2q
\end{equation}

\begin{equation}
\vec{E}^{\parallel}=\int{\frac{i\pi\omega^{SiO_{2}}{t_{10}^{\parallel}}}{{\sqrt{\varepsilon_{SiO_{2}}}}q^2 s_{1}}}exp[i\vec{q}.\vec{\rho}+ipx]\times[q^2 g_{x}(\vec{k}) -s_{1}(\vec{q}.\vec{g}(\vec{k}))](q^2\hat x-p\vec{q}]d^2q
\label{eq:wideeq}
\end{equation}
\end{widetext}
where,$\vec{E}^{\perp}$ and $\vec{E}^{\parallel}$ arise from the wave polarized perpendicular and parallel to the plane of incidence, respectively and $t_{10}^{\perp}$  and $t_{10}^{\parallel}$  are the transmission coefficients when the electric field is perpendicular and parallel to the plane of incidence, respectively. 
     It can be shown that, if the vacuum width $d$ satisfies in the below equation:
\begin{equation}
d\ll\frac{c}{2\omega \sqrt{\mid(1-\varepsilon_{1,2})\mid}} 
\end{equation}
where, $\varepsilon_{1,2}=\varepsilon_{SiO_{2}}$ or $\varepsilon_{SG}$. The evanescent part of the heat flow, $P_{ev}$, is equal to\cite{R5}:
\begin{widetext}
\begin{equation}
P_{ev}={\frac{\hbar}{\pi^2 d^2}}\int_{0}^{\infty}\omega d\omega(n_{B}(E,T_{1})-n_{B}(E,T_{2}))\times\int_{0}^{\infty}\gamma d\gamma {\frac{Im(\varepsilon_{1}) Im(\varepsilon_{2})}{(\mid(\varepsilon_{1}+1)(\varepsilon_{2}+1)-(\varepsilon_{1}-1)(\varepsilon_{2}-1) e^{-\gamma} \mid)^2}}e^{-\gamma}
\label{eq:wideeq}
\end{equation}
\end{widetext}

Where, $\gamma=-2p\omega d/c$ and  $n_{B} (E,T_{i})=[e^{β\hbar\omega}-1]^{-1}$. If the SG layer is absent, then:
\begin{widetext}
\begin{equation}
X\equiv\int_{0}^{\infty}\gamma d\gamma {\frac{Im(\varepsilon_{1}) Im(\varepsilon_{2})}{(\mid(\varepsilon_{1}+1)(\varepsilon_{2}+1)-(\varepsilon_{1}-1)(\varepsilon_{2}-1) e^{-\gamma} \mid)^2}}e^{-\gamma}=\int_{0}^{\infty}\gamma d\gamma {\frac{Im(\varepsilon_{1}) Im(\varepsilon_{2})}{(\mid(\varepsilon_{G/SiO_{2}}+1)^2-(\varepsilon_{G/SiO_{2}}-1)^2 e^{-\gamma} \mid)^2}}e^{-\gamma}
\label{eq:wideeq}
\end{equation}
\end{widetext}
$X$ is radiative heat flux in the near-field and after normalization is shown by$ X^*$\cite{R14}. In next section, we will use Eqs.20, 21, 32 and 33 for calculating $P_{ev}$ .
\section{Numerical calculations and discussion}
As an example, let us to consider two materials both with $-5\leq Re(\varepsilon)\leq 5$ and $0.001\leq Im(\varepsilon)\leq 10$ and study the heat flux between them. Fig.2(a) and (b) shows the normalized heat flux $X^*$  when the thickness of gap $d$ is equal to zero and 10 nm, respectively. 

\begin{figure}[b]
\includegraphics{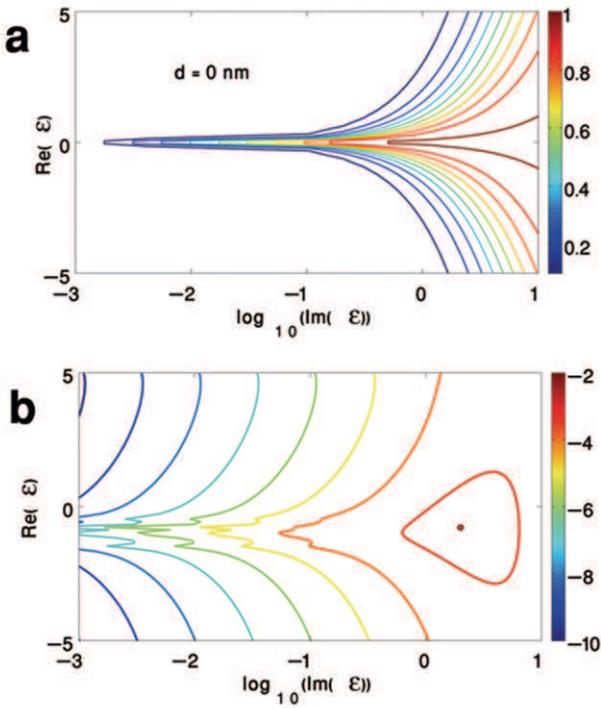}% Here is how to import EPS art
\caption{\label{fig:epsart}  (Color online) (a) Contour plot of $X^*$as a function of $Re(\varepsilon)$ and $Im(\varepsilon)$ at $d=0 nm$ and (b) $log_{10} (X^*)$  at $d=10 nm$.}
\end{figure}

As Fig.2 shows, only for specific values of $Re(\varepsilon)$ and $Im(\varepsilon)$ heat flux is equal to one at $d=0$ nm and decreases to the maximum value 0.01 at $d=10$ nm due to the dependency of $X^*$ to $d$. The result is in good agreement with the result of Ref.14.
Now, we consider two $G/SiO_{2}$ layers and calculate the heat flux between them. Since, $k_{B}=1.381\times 10^{-23} m^2 Kg s^{-2} K^{-1}$, $h=6.62\times 10^{-34} m^2 Kg s^{-1}$, $c=3\times 10^8 m s^{-1}$, and $T=300 K$, therefore, $\omega=k_{B} T/\hbar=3.932\times 10^{13} s^{-1}$. Also, we know $\varepsilon_{SiO_{2}}=4$\cite{R23} then, by using Eqs. 20 and 21 we find:
\begin{equation}
 Re(\varepsilon_{G/SiO_{2}} )=1.125 
\end{equation}
 \begin{equation}
Im(\varepsilon_{G/SiO_{2}} )={\frac{e^2}{32}}{\frac{x}{\sqrt{v^2-x^2 }}}                                                                                   
\end{equation}
If we assume, for example $\varepsilon_{G/SiO_{2}}=1.125+i0.05$ then, the Eq.31 is satisfied for $d\ll28.3\times 10^{-6} m$. We assume $0\leq d \leq 10 nm$ in next calculations.
 
Fig.3 shows contour plot of $log_{10}{X}^*$ as a function of $Im(\varepsilon_{G/SiO_{2}})$ and $d$. Here, we assume  $x=q/k_{F}=1$ , and $k_{B}=\hbar=e=1$, for simplicity. 

\begin{figure}[b]
\includegraphics{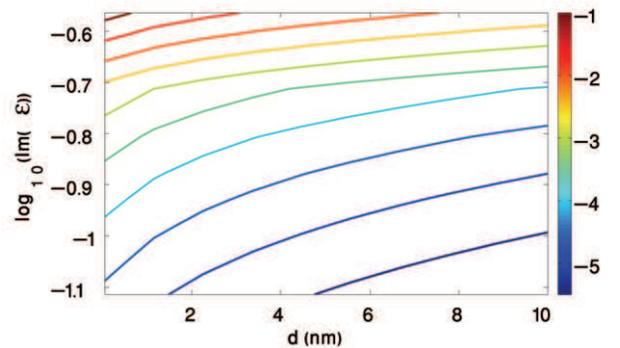}% Here is how to import EPS art
\caption{\label{fig:epsart}(Color online) Contour plot of $log_{10}X^*$as a function of $Im(\varepsilon)$ and $d=0 nm$.}
\end{figure}

According to Fig.3, the heat flux reaches to 0.1 when imaginary part of dielectric constant is at range $0.20<Im(\varepsilon)\leq 0.27$ and $0\leq d \leq7 nm$. In comparison with Fig.2, the heat flux increases by one order of magnitude. It means that the heat flux is amplified by choosing the $G/SiO_{2}$ structure.By using the value of $Im(\varepsilon)$ and Eq.28 we find $1.006<v=\omega/E_{F}\leq1.012$.  On the Plasmon mode dispersion curve, the point $(q/k_{F},\omega/E_{F} )=(1,1.01)$ is placed on the boundaries of the single particle excitation (SPE) for intra- and inter-band electron excitation in graphene \cite{R31, R32}. Thus, the heat flux has a maximum at Plasmon frequency supported by the $G/SiO_{2}$ slabs.

Now let us to add the SG layer between two $G/SiO_{2}$ layers as Fig.1 shows. The total heat flow on SG layer from left $G/SiO_{2}$ layer is proportional to $[n_{B} (E,T_{1} )-n_{B} (E,T_{2} )]$ and the total heat flow on right $G/SiO_{2}$ layer from SG layer is equal to $[n_{B} (E,T_{2} )-n_{B} (E,T_{3} )]$. We choose the temperature $T_{2}$ the value such that the total heat flux on SG layer is zero i.e.,
\begin{equation}
n_{B} (E,T_{1} )-n_{B} (E,T_{2} )=n_{B} (E,T_{2} )-n_{B} (E,T_{3})  
\end{equation}                              
or\\
\begin{equation}
T_{2}={\frac{1}{k_{B}\hbar\omega}} Ln[1+{\frac{2(e^{\beta_{1}\hbar\omega}-1)( e^{\beta_{3}\hbar\omega}-1)}{(e^{\beta_{1}\hbar\omega}-1)+( e^{\beta_{3}\hbar\omega}-1)}}]                                                  
\end{equation}  
\\
Here, we assume $T_{1}=300 K$, $T_{3}=323 K$.
Fig.4 (a) shows the heat flux $log_{10} {[n_{B} (E,T_{i} )-n_{B} (E,T_{j} )] X^*}$ between right (left) $G/SiO_{2}$ layer and SG layer and Fig.4 (b) shows the heat flux between two $G/SiO_{2}$ layers.
\\
\begin{figure}[b]
\includegraphics{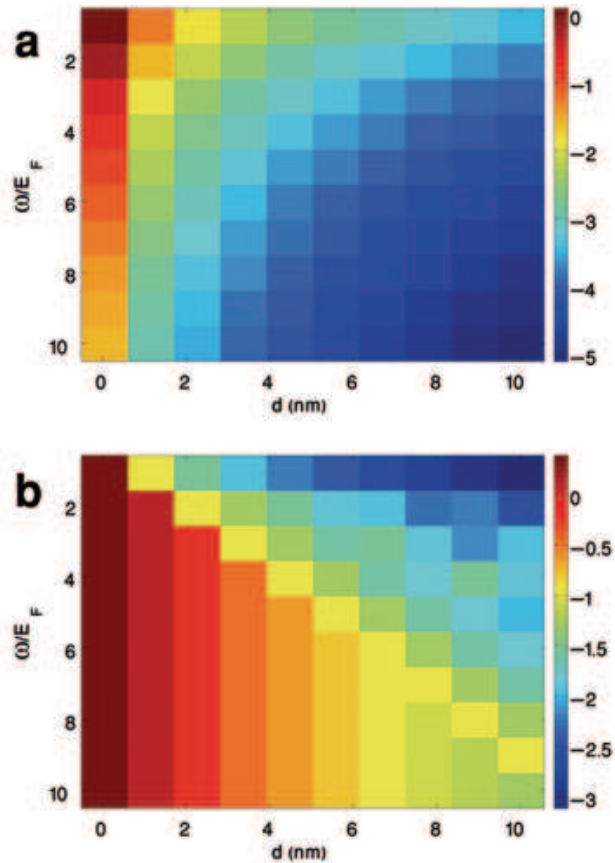}% Here is how to import EPS art
\caption{\label{fig:epsart}(Color online) (a) Heat flux between $G/SiO_{2}$ layer and SG layer and (b) between two $G/SiO_{2}$ layers. Here, colorbar shows $log_{10} {[n_{B} (E,T_{i} )-n_{B} (E,T_{j} )] X^*}$ .}
\end{figure}
The heat flow between $G/SiO_{2}$ and SG layer is maximized for $d \rightarrow 0$ and $\omega/E_{F} \rightarrow 1$. Therefore, it is maximized at Plasmon frequency which is placed on the boundaries of the SPE for intra- and inter-band electron excitation (note, $x={q}/{k_{F}} =1$). As Fig.4 (b) shows, for $1\leq {\omega}/{E_{F}} \leq 10$ and $d\leq 1 nm$, then $[n_{B} (E,T_{i} )-n_{B} (E,T_{j} )] X^* \rightarrow 0.5$. Also for $2\leq {\omega}/{E_{F}} \leq 10$ and $d\leq 3 nm$, then $-0.5<[n_{B} (E,T_{i} )-n_{B} (E,T_{j} )] X^*<0.5$. It means that for a wide range of Plasmon frequency and gap thickness the heat flux has maximum. 
Finally, as Fig.5 shows, the ratio of heat flux after adding the SG layer to before adding the layer depends on $\omega/E_{F}$ and $d$ but the heat flux is only amplified for specific range of $\omega/E_{F}$ and $d$ by adding the intermediate SG layer. 
\\
\begin{figure}[b]
\includegraphics{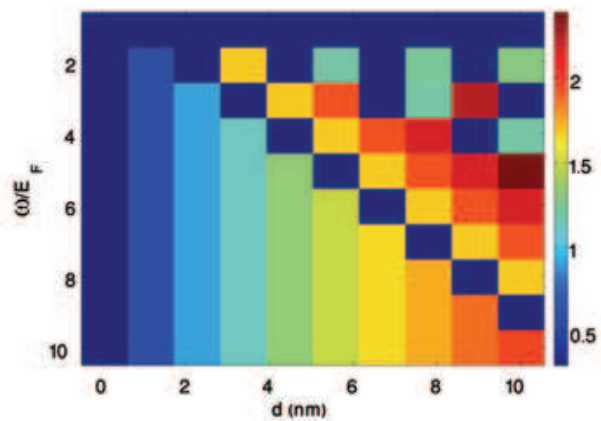}% Here is how to import EPS art
\caption{\label{fig:epsart}(Color online) The ratio of heat flux after adding the SG layer to before adding the layer.}
\end{figure}
\\
It can be understood by looking at the coupling of modes between each $G/SiO_{2}$ layer and middle SG layer \cite{R23}. i.e., since the total heat transmission between two $G/SiO_{2}$ layers $\Phi_{13}$ is proportional to:
\begin{equation}
\Phi_{13}\propto [n_{B} (E,T_{1} )-n_{B} (E,T_{2} )] X^*+[n_{B} (E,T_{2} )-n_{B} (E,T_{3} )] X^*   
\end{equation} 
 and  $n_{B,12}=n_{B,23}=n_{B,13}/2$, therefore $\tau_{13}=(\tau_{12}+\tau_{23})/2$ where $\tau_{ij}$ is the 
transmission probability between layer $i$ and layer $j$ \cite{R23}. 
\\
\section{Summary}
We have considered a suspended graphene (SG) layer between two Graphene on $SiO_{2}$ layers $(G/SiO_{2})$ and studied the heat flux amplification between two $G/SiO_{2}$ layers. It has been shown that, before adding the SG layer, the heat flux had maximum at Plasmon frequency supported by the $G/SiO_{2}$ slabs. By adding the SG layer, the heat flux between two $G/SiO_{2}$ was amplified, for specific range of vacuum gap between SG layer and $G/SiO_{2}$ layers and Plasmon freqency, due to the modes coupling between each $G/SiO_{2}$ layer and middle SG layer. Since the intermediate SG layer was a single atmoic layer, the heat transfer did not depend on the thickness of middle layer. 
\nocite{*}

\bibliography{apssamp}% Produces the bibliography via BibTeX.

\end{document}